\documentstyle[aps,multicol,prb,epsf]{revtex} 
\begin{document} 
%
% Defintions 
%
\newcommand{\be}{\begin{equation}}
\newcommand{\ee}{\end{equation}}
\newcommand{\bea}{\begin{eqnarray}}
\newcommand{\eea}{\end{eqnarray}}
\newcommand{\beann}{\begin{eqnarray*}}
\newcommand{\eeann}{\end{eqnarray*}}
\newcommand{\bma}{\begin{array}{cc}}
\newcommand{\ema}{\end{array}}
\newcommand{\fr}{\frac}
\newcommand{\ra}{\rangle}
\newcommand{\la}{\langle}
\newcommand{\li}{\left}
\newcommand{\re}{\right}
\newcommand{\ri}{\right}
\newcommand{\uarr}{\uparrow}
\newcommand{\darr}{\downarrow}
\newcommand{\alp}{\alpha}
\newcommand{\df}{\stackrel{\rm def}{=}}
\newcommand{\nn}{\nonumber}
\newcommand{\dpl}{\displaystyle}
\def \gplus#1{{\cal G}^+ \left( #1 \right) }
\def \gminus#1{{\cal G}^- \left( #1 \right) }
\def \diff#1{{\cal D} \left( #1 \right) }
\def \lp {L_{\phi}}
\def \xo {x_1}
\def \xop {x_1^{\prime}}
\def \xt {x_2}
\def \yo {y_1}
\def \yop {y_1^{\prime}}
\def \yt {y_2}
\draft
\title{Ballistic Quantum Transport: Effect of Geometrical Phases}
\author{Diego Frustaglia and Klaus Richter} 
\address{Max-Planck-Institut f\"{u}r Physik komplexer Systeme,
N\"othnitzer Str. 38, 01187~Dresden, Germany} 
\date{September 5, 2000}
%\date{\today} 
\maketitle
\begin{abstract} 
{
We study the influence of nonuniform magnetic fields on the magneto conductance
of mesoscopic microstructures. We show that the coupling of the electron spin
to the inhomogenous field gives rise to effects of the Berry phase on ballistic quantum 
transport and discuss adiabaticity conditions required to observe such effects.
We present numerical results for different ring geometries showing a splitting
of Aharonov-Bohm conductance peaks for single rings and corresponding signatures 
of the geometrical phase in weak localization. The latter features can be
qualitatively explained in a semiclassical approach to quantum transport.
}
\end{abstract} 
%03.65.Bz Foundations, theory of measurement, miscellaneous theories 
%   (including Aharonov- Bohm effect, Bell inequalities, Berry's phase)
%03.65.Sq Semiclassical theory and applications
%05.45.+b theory and model of chaotic systems
%05.30.Fk Fermion system and electron gas
%73.20.Dx Electron states in low dimensional structures
%
%\ifpreprintsty\tightenlines \else \begin{multicols}{2} \fi
\bibliographystyle{simpl1}
%\begin{multicols}{2}
%
%%%%%%%%%%%%%%%%%%%%%%%%%%%%%%%%%%%%%%%%%%%%%%%%%%%%%%%%%%%%%%%%%%%%
%%%%%%                  NEW SECTION
%%%%%%%%%%%%%%%%%%%%%%%%%%%%%%%%%%%%%%%%%%%%%%%%%%%%%%%%%%%%%%%%%%%%
%
\section{Introduction}
Quantum transport through ballistic conductors,
mesoscopic systems of reduced dimensionality where impurity
scattering is strongly reduced, has been intensively studied 
throughout the last decade\cite{ferry}. 
In these phase-coherent ballistic cavities, built from high-mobility semiconductor 
heterostructures, characteristic quantum phenomena in the conductance
have been observed such as Aharonov-Bohm oscillations, conductance fluctuations, 
or weak localization. Such  quantum transport effects, which were originally 
known from disordered systems, have been related to interference  
of electron waves which undergo multiple reflections at the boundaries of
the confining potentials. They represent features of the 
{\em orbital} dynamics of electrons in confined systems. 

Here we consider {\em spin} effects on ballistic quantum transport.
In particular in the recent past
the r\^ole of spin in mesoscopic devices, sometimes referred to as
`spintronics', has received considerable attention, for instance in 
the context of Coulomb blockade, spin injection mechanisms, or 
quantum computing, to name only a few directions. 
We focus on spin effects due to
geometrical phases which may arise, besides e.g.\ Aharonov-Bohm phases\cite{aharonov},
from the coupling of the electron spin to nonuniform fields on mesoscopic 
scales. If the electron spin can adiabatically follow a spatially
varying magnetic field the spin wave function acquires a
geometrical or Berry phase\cite{berry}. 
In the context of mesoscopic physics, Berry phases were first studied 
for one-dimensional (1d) rings 
by Loss, Goldbart and Balatsky\cite{loss}, and by Stern\cite{stern}.
In later work  the r\^ole of geometrical phases on 
weak localization and conductance fluctuations in disordered
samples was addressed \cite{loss2} and the Berry phase effect on persistent
currents was reconsidered\cite{kawabata}. 
The question whether the elastic scattering time 
or the Thouless time sets the relevant time scale for adiabaticity in a 
disordered conductor 
has recently led to a controversial
discussion\cite{beenakker,loss3} 
Related geometrical phases arising
from spin-orbit interaction  have also been extensively studied in the
literature\cite{spin-orbit}.  

\begin{figure}
\begin{center}
\leavevmode
\epsfxsize=0.35\textwidth
\epsfbox{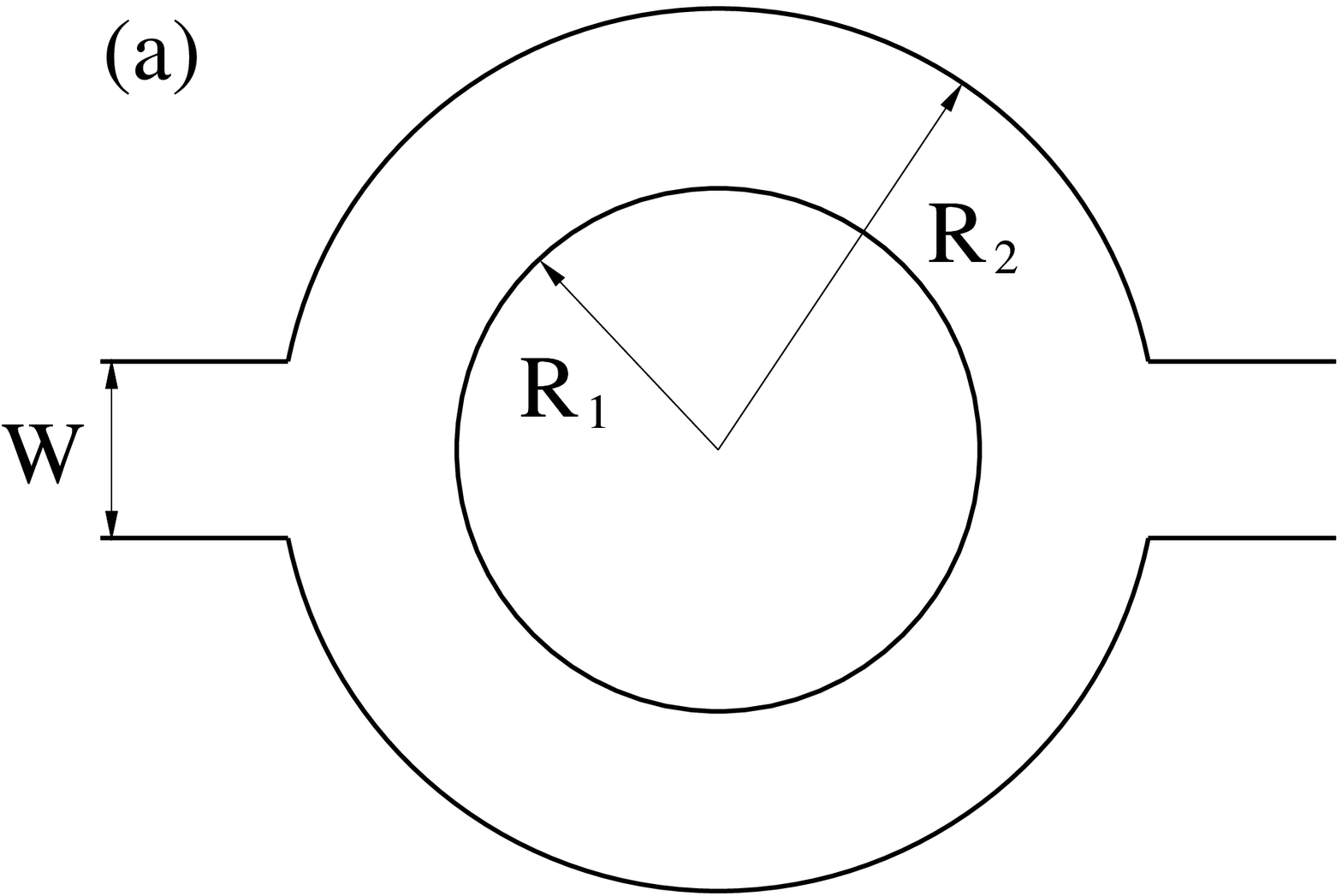}
\hspace*{2cm} 
% \vspace*{1cm}
\leavevmode
\epsfxsize=0.40\textwidth
\epsfbox{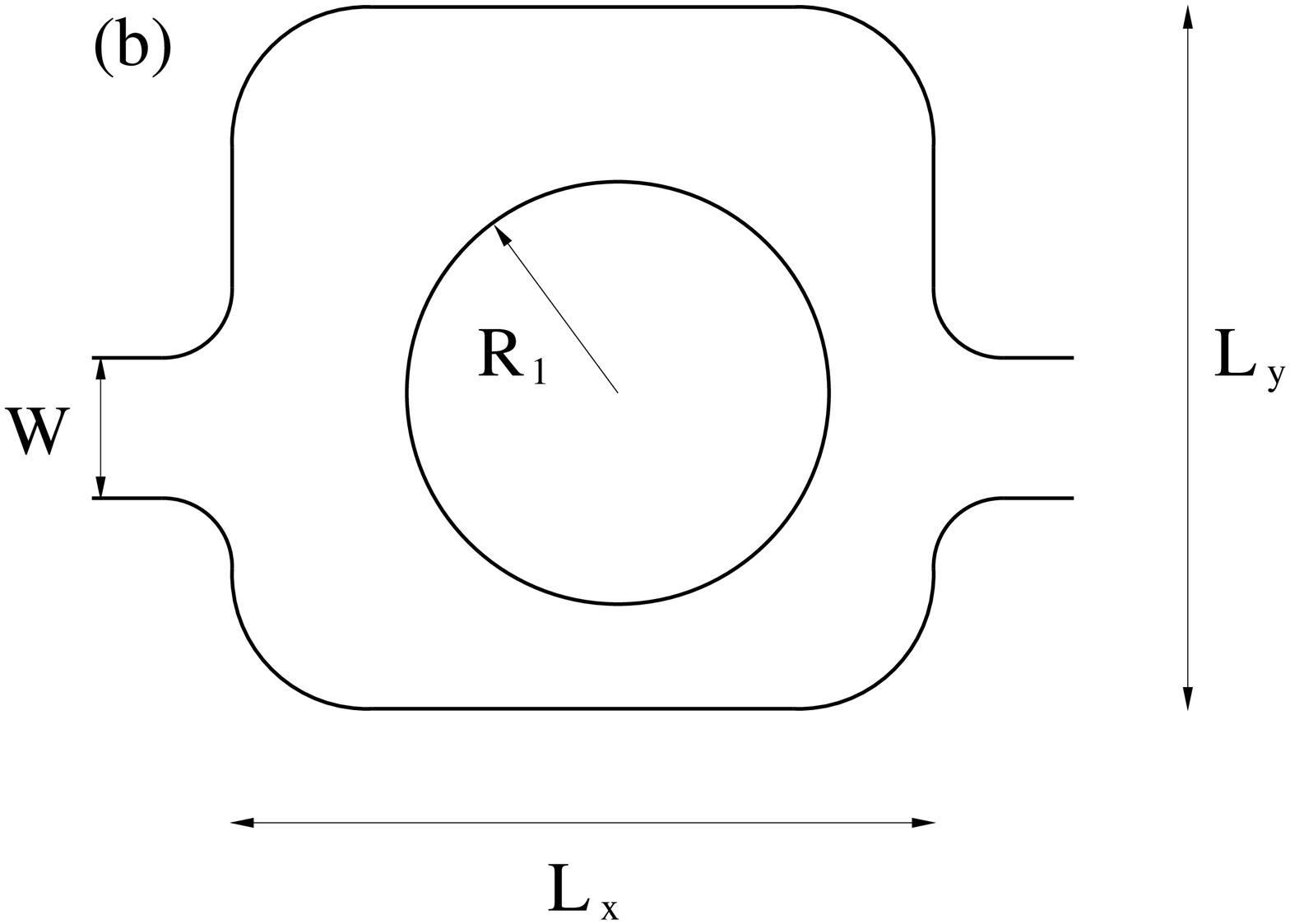}
\end{center}
\caption{
Geometries of ballistic microstructures used in the numerical 
quantum calculations of the conductance. 
}
\label{fig:model}
\end{figure}

We consider Berry phase effects on the conductance through
{\em ballistic} Aharonov-Bohm (AB) rings or, more generally, two-dimensional (2d)
doubly-connected structures 
with topologies such as  shown in Fig.~\ref{fig:model}. 
Besides a weak uniform magnetic probe field we assume the presence of a 
nonuniform magnetic field, which is rotationally symmetric with respect 
to the antidot close to the center of the structure (see Fig.\ \ref{fig:Bfield}).

In a novel experimental approach to observe the Berry phase
such a setup has been recently realized for a two-dimensional electron gas
(2DEG) by placing a micromagnet at the center of a ring geometry fabricated
from a high-mobility GaAs-AlGaAs heterostructure\cite{weiss}.
The micromagnet creates a nonuniform field with a tilt solid angle 
which varies on mesoscopic length scales. The measured magneto 
conductance showed clear signatures of the inhomogenous field, 
e.g.\ beating patterns in the AB oscillations. Berry phase effects have been
proposed as one possible mechanism to describe these features.
A related experiment showed a splitting of the $h/e$-peak in the Fourier 
transform of AB conductance oscillations\cite{morpurgo}. 
This splitting was ascribed to a geometrical phase owing to strong 
Rashba spin-orbit interaction in the InAs samples used.  
For a most recent experiment on Rashba splitting see, e.g.\ Ref.~\cite{grundler}.
 
The above mentioned ballistic microstructures can be viewed as electron
quantum billiards. Hence they allow for studying signatures of the
classical (electron) dynamics in quantum properties such as the
conductance\cite{baranger1}. The link between classical motion, 
which usually is nonintegrable, and quantum dynamics can be provided by 
semiclassics\cite{reviews}. Here we outline a semiclassical approach 
in order to understand Berry phase effects in chaotic quantum transport.
The case of a geometrical phase has also been considered in
semiclassical approaches to multicomponent wave fields\cite{LF91}
and the Dirac equation\cite{BK98}.

We first descibe in Sec.\ \ref{sec:hamiltonian}
the adiabatic treatment leading to the geometrical phase.
We extend an earlier approach\cite{stern} to mesoscopic rings of
finite width and discuss conditions to achieve adiabaticity 
in high-mobility heterostructures. In Sec.\ \ref{sec:approach} we 
summarize our quantum mechanical approach to compute
the conductance. In Sec.\ \ref{sec:results} we
present numerical results for AB rings
 and generalizations to asymmetric geometries as the
example shown in Fig.\ \ref{fig:model}(b). We first show that
single phase-coherent rings, both with one and more transverse channels,
exhibit splittings in the  peaks of 
magneto conductance oscillations resulting from the different
phase acquired by electrons with different spin 
quantum number. 
We further find clear signatures of the Berry phase on the 
average resistance, i.e.\ weak localization. The numerical
quantum results can be qualitatively understood in a semiclasscal 
picture invoking chaotic electron transport in nonuniform
magnetic fields.

%
%%%%%%%%%%%%%%%%%%%%%%%%%%%%%%%%%%%%%%%%%%%%%%%%%%%%%%%%%%%%%%%%%%%%
%%%%%%                  NEW SECTION
%%%%%%%%%%%%%%%%%%%%%%%%%%%%%%%%%%%%%%%%%%%%%%%%%%%%%%%%%%%%%%%%%%%%
%

\section{Hamiltonian and Berry Phase}
\label{sec:hamiltonian}

The concept of a geometrical phase follows from an adiabatic treatment of the
Schr\"odinger equation for the spin wave function. 
We first decompose the Hamiltonian for electrons in ring geometries into an
adiabatic part, diagonal in the spin quantum number,
and nonadiabatic corrections which couple different spin degrees of freedom.
We then discuss the conditions under which adiabaticity is achieved.

\subsection{Adiabatic approach}

We consider noninteracting electrons with charge $-e$, effective mass
$m^\ast$ (bare mass $m_0$), and spin described by the Pauli spin matrix
vector $\vec{\sigma}$. In the presence of a magnetic field
$\vec{B}=\vec{\nabla} \times \vec{A}_{\rm em}$ the Hamiltonian reads
\be
\label{qham}
H =  \fr{1}{2m^*}\li[\vec{p}+
    \fr{e}{c}\vec{A}_{\rm em}(\vec{r})\re]^2+ V(\vec{r}) +
    \mu\ \vec{B} \cdot\vec{\sigma} \; ,
\ee
with magnetic moment $\mu= g^* \mu_{\rm B}/2 =  g^* e \hbar/(4m_0\ c)$ and effective 
gyromagnetic ratio $g^*$. The electrostatic potential $V(\vec{r})$
defines for instance the confining potential of a 2d ballistic conductor. 
In the case we are interested in, the electromagnetic vector potential has two 
contributions, $\vec{A}_{\rm em}= \vec{A}_0+\vec{A}_{\rm i}$. The term
$\vec{A}_0$ generates a uniform magnetic field $\vec{B}_0(\vec{r}) = B_0\hat{z}$, 
which points perpendicular to the plane of the electron gas and which is the tunable
parameter used to study the magneto conductance of the system. The term
$\vec{A}_{\rm i}(\vec{r})$ represents an inhomogenous magnetic field 
$\vec{B}_{\rm i}(\vec{r})$. Fig.~\ref{fig:Bfield}(a) shows 
an example chosen for our numerical analysis given below.

In principle, if $\mu B$ is sufficiently large, we can separate the
Hamiltonian into an adiabatic part  and nonadiabatic corrections.
Following Refs.\ \cite{stern,aharonov2}, we
start by defining a basis $\cal{B}$ of spin eigenstates of the Zeeman 
term in Eq.\ (\ref{qham}), 
${\cal B}(\vec{r})=\li\{ |\uarr(\vec{r}) \ra ;|\darr(\vec{r}) \ra \re\}$.
Here, $|\uarr(\vec{r}) \ra$ and $|\darr(\vec{r}) \ra$ represent 
spin states parallel and antiparallel to the local direction 
$\hat{n}(\vec{r})=\vec{B}(\vec{r})/B(\vec{r})$ of the magnetic field.
The adiabatic approximation consists in treating $\vec{r}$ as (slowly varying)
parameter. 

Projecting the full Hamiltonian $H$ of orbital and spin degrees of freedom
onto the defined subspaces of spin-up and -down states, given by the operators
${\cal P}^{\uarr(\darr)}=\fr{1}{2}(1\pm\hat{n}\cdot\vec{\sigma})$, 
enables one to decompose $H$ into adiabatic and nonadiabatic contributions
\be
\label{qhamp}
H=H_{\rm d}+H_{\rm nd} \; .
\ee 
Here, the diagonal part
$H_{\rm d}={\cal
P}^{\uarr}H{\cal P}^{\uarr}+{\cal P}^{\darr}H{\cal P}^{\darr}$ consists of
matrix elements which are non-zero only within each spin subspace.
The matrix elements of the nondiagonal term,
$H_{\rm nd}={\cal P}^{\uarr}H{\cal P}^{\darr}+{\cal P}^{\darr}H{\cal
P}^{\uarr}$, are non-zero only when taken between
different subspaces.
It was shown that by defining 
\be
\vec{A}=\vec{p}- {\cal P}^\uarr \vec{p}\ {\cal P}^\uarr - 
 {\cal P}^\darr \vec{p}\ {\cal P}^\darr 
\ee
one can express $H_{\rm d}$ and $H_{\rm nd}$ as\cite{stern,aharonov2}
\bea
\label{qhamd}
H_{\rm d} & = & \fr{1}{2m^*}\li[ (\vec{\Pi}-\vec{A})^2+\vec{A}^2
\re]+V+\mu\ \vec{B}\cdot\vec{\sigma} \; , \\
\label{qhamnd}
H_{\rm nd} & = & \fr{1}{2m^*}\li[(\vec{\Pi}-
\vec{A})\cdot\vec{A}+\vec{A}\cdot(\vec{\Pi}-\vec{A})\re] 
\eea
with $\vec{\Pi}=\vec{p}+(e/c) \vec{A}_{\rm em}$.
Using the definitions given above we can also express $\vec{A}$ as
$\vec{A}=(i\hbar/2)(\hat{n}\cdot\vec{\sigma})
\vec{\nabla}(\hat{n}\cdot\vec{\sigma})$. 

By construction $H_{\rm d}$ has a set of eigenstates $|N,\uarr \ra =
\psi^\uarr_N(\vec{r}) \otimes |\uarr(\vec{r}) \ra$ with spin parallel to the
field and $|N,\darr \ra = \psi^\darr_N(\vec{r}) \otimes |\darr(\vec{r}) \ra$
with spin antiparallel to the field. $\psi^{\uarr (\darr)}_N(\vec{r})$ denote the
corresponding spatial wave functions and the index $N$ indicates the set
of spatial quantum numbers. 
% (e.g. mode and/or orbital quantum numbers and so on). 
In the following we omit this index. Defining a unitary operator ${\cal
U}(\vec{r})$ such that ${\cal U}|\uarr(\vec{r}) \ra={1 \choose 0}$ and ${\cal
U}|\darr(\vec{r}) \ra={0 \choose 1}$, we diagonalize $H_{\rm d}$ and find 
as an effective Hamiltonian for the orbital motion 
(with eigenfunctions $\psi^{\uarr (\darr)}(\vec{r})$)
\be
\label{qhamdB-m}
{\cal U}\ H_{\rm d}\ {\cal U}^{\dag}=
\left( \begin{array}{cc}
H^\uarr & 0 \\
0 & H^\darr \\
\end{array} \right)
\ee
with
\be
\label{qhamdB}
H^{\uarr(\darr)}=
\fr {1}{2m^*}{\Big [} \vec{\Pi}-\vec{A}_{\rm g}^{\uarr(\darr)} 
{\Big ]}^2 +  V + V_{\rm eff}^{\uarr(\darr)} \; .
\ee
In Eq.~(\ref{qhamdB}),
$\vec{A}_{\rm g}^{\uarr(\darr)}$ and $V_{\rm eff}^{\uarr(\darr)}$ are the 
geometrical vector potential and an effective  potential,
respectively, which arise from the projection. In cylinder
coordinates $\vec{r} = r \ \hat{r}+ z \ \hat{z}$ with $\hat{r}=\cos \varphi \
\hat{x}+\sin \varphi \ \hat{y}$ and $\hat{\varphi}=-\sin \varphi \ \hat{x}+\cos
\varphi \ \hat{y}$,
% where $r\hat{r}$ is the projection of $\vec{r}$ onto the $xy$-plane,
they take the form 
\bea
\label{Ag-gen}
\vec{A}_{\rm g}^{\uarr(\darr)}(\vec{r}) & = & 
\fr{\hbar}{2} \li[1\pm \cos\alp(\vec{r}) 
\re]\vec{\nabla}[\varphi+\varphi_0(\vec{r})] \; , \\
\label{Veff-gen}
V_{\rm eff}^{\uarr(\darr)}(\vec{r}) & = & 
\fr{\hbar^2}{8m^*}\li\{\sin^2\alp(\vec{r}) 
\li[\vec{\nabla}[\varphi+\varphi_0(\vec{r})] \re]^2+
\li[\vec{\nabla}\alp(\vec{r}) \re]^2  \re\} \pm \mu B(\vec{r}) \; .
\eea
Here, $\alp$ is the angle between the z-axis and the local direction of the field $\vec{B}$,
and $\varphi_0$ is the polar angle between the projections of $\vec{B}$ and $\vec{r}$
onto the x-y plane. If the
magnetic field has the symmetry $\vec{B}(\vec{r})=\vec{B}(r)$, and if
$\varphi_0$ is independent of $r$, 
 i.e. $\vec{\nabla}(\varphi+\varphi_0)=(1/r)
\hat{\varphi}$, Eqs.\ (\ref{Ag-gen}) and (\ref{Veff-gen}) are reduced to
the form 
\bea
\label{Ag}
\vec{A}_{\rm g}^{\uarr(\darr)}(\vec{r}) & = & 
\fr{\hbar}{2r} \li[1\pm \cos\alp(r) \re]\hat{\varphi} \; , \\
\label{Veff}
V_{\rm eff}^{\uarr(\darr)}(\vec{r}) & = & 
\fr{\hbar^2}{8m^*}\li[\fr{1}{r^2}\sin^2\alp(r) +
\li(\fr{\partial \alp}{\partial r} \re)^2  \re] \pm \mu B(r) \; .
\eea
We note that $V$ need not obey any symmetry properties.

Correspondingly, we find from $H_{\rm nd}$ the effective Hamiltonian
\be
\label{qhamndB-m}
{\cal U}\ H_{\rm nd}\ {\cal U}^{\dag}=
\left( \begin{array}{cc}
0 & H^{\uarr \darr} \\
H^{\darr \uarr} & 0 \\
\end{array} \right)
\ee
with
\bea
\label{qhamndB}
H^{\uarr \darr (\darr \uarr)}= 
\fr{1}{2m^*} \Bigg{\{} \fr {\hbar}{2}\li[ \sin \alp(\vec{r})
\vec{\nabla}[\varphi+\varphi_0(\vec{r})] \pm {\rm i} \vec{\nabla}\alp(\vec{r}) \re]
\cdot \li[2 \vec{\Pi}-\hbar \vec{\nabla}[\varphi+\varphi_0(\vec{r})] \re]-\nn 
\\
-\fr {{\rm i} \hbar^2}{2} \Big{\{} \vec{\nabla}\cdot \li[\sin \alp(\vec{r})
\vec{\nabla}[\varphi+\varphi_0(\vec{r})]\pm {\rm i} \vec{\nabla}\alp(\vec{r})
\re] \Big{\}}
\Bigg{\}} \; .
\eea   
If $\vec{B}(\vec{r})=\vec{B}(r)$ with constant $\varphi_0$ and
if we choose a proper gauge, we can write $\vec{A}_{\rm
em}=A_\varphi \hat{\varphi}+A_{\rm z} \hat{z}$. Hence Eq. (\ref{qhamndB}) can
be reduced to  
\be
\label{qhamndB-sym}
H^{\uarr \darr (\darr \uarr)}= \fr {\hbar^2}{2m^*} {\Bigg \{} - \fr {1}{2r} \sin
\alp(r) \li[-\fr {2e}{\hbar c} A_\varphi(r) + \fr {1}{r} + \fr {2{\rm i}}{r} \ \fr
{\partial}{\partial \varphi} \re] \pm \li[ \fr {1}{2} \li( \fr {1}{r} \ \fr
{\partial \alp}{\partial r} + \fr {\partial^2 \alp}{\partial r^2} \re) + \fr
{\partial \alp}{\partial r} \ \fr {\partial}{\partial r}\re] {\Bigg \}} \; . 
\ee   

The adiabatic approximation consists in neglecting $H_{\rm nd}$ in
Eq.\ (\ref{qhamp}), i.e.\  neglecting the off-diagonal contribution
(\ref{qhamndB-m},\ref{qhamndB}).  
The term proportional to $\sin \alp$ in Eq.\ (\ref{qhamndB-sym}) 
corresponds to the case of a 1d ring\cite{stern} with fixed $r$. 
The second term in Eq.~(\ref{qhamndB-sym}) is related to radial motion and
leads to additional adiabaticity conditions which have to be fulfilled,
as will be discussed below.

In the adiabatic approximation $H \simeq H_{\rm d}$ the system decouples
into two (independent) electron gases described by the effective Hamiltonians
$H^{\uarr (\darr)}$ in Eq.\ (\ref{qhamdB}). They characteristically differ
in the contribution of the geometrical vector potential to the kinetic energy
terms giving rise to an effective vector potential 
\be
\label{Aeff}
\vec{A}_{\rm eff}^{\uarr(\darr)}=\vec{A}_{\rm em}-\fr{c}{e}\vec{A}^{\uarr(\darr)}_{\rm g}\; .
\ee
The corresponding effective flux enclosed by a closed path $\Gamma$ is 
\be
\label{phi-eff}
\phi_{\rm eff}^{\uarr(\darr)}=\oint_{\Gamma}\vec{A}_{\rm eff}^{\uarr(\darr)} \cdot 
\vec{{\rm d}l}=\phi+\phi_{\rm i}-\phi_0 \  \fr{\gamma^{\uarr(\darr)}}{2\pi} \; .
\ee
Here, $\phi={\cal A}_{\Gamma} B_0$ is the flux of the uniform field
through the enclosed area ${\cal A}_{\Gamma}$, and $\phi_{\rm i}$ is the
contribution from the nonuniform field. $\phi_0 = hc/e$ is the flux
quantum. For $\vec{A}^{\uarr(\darr)}_{g}$ as given in Eq. (\ref{Ag}) 
the Berry phase $\gamma^{\uarr(\darr)}$ takes the form
\be
\label{Bphase}
\gamma^{\uarr(\darr)}=\oint_{\Gamma}\fr{1}{2r} \li[1\pm \cos\alp(r)
\re]\hat{\varphi} \cdot \vec{{\rm d}l} \; .
\ee
Eq.\ (\ref{Bphase}) shows that during one round trip 
$\gamma^{\uarr(\darr)}$ can
vary only in the range $[0,2\pi]$ (since $0 \le \alp \le \pi$)
limiting its contribution to one flux quantum at most. This
property distinguishes the geometrical from the electromagnetic flux. 
% All those contributions depend on the particular path $\Gamma$.

% \section{Adiabatic approximation in two-dimensional (2d) ballistic rings}

\subsection{Conditions for adiabaticity in two-dimensional ballistic rings}

In the following we discuss under which conditions an adiabatic treatment
of ballistic transport in nonuniform fields is justified, and we 
give some implications for possible experimental observations of the Berry phase.

In the adiabatic limit, $\vec{r}$ is treated as a parameter in $H(\vec{r})$
when diagonalizing the spin dependent part of the Hamiltonian at every point
in space.
Adiabaticity is achieved if the electron motion is slow enough such that the
magnetic moment associated with the spin stays (anti)aligned with the
local inhomogenous magnetic field. 
This requires a separation of time scales: The Larmor frequency of
spin precession,  $\omega_{\rm s}=2 \mu B/\hbar$, must be large compared 
to the inverse time it takes the electron to traverse a distance over
which the direction of the field $\hat{n}$ changes significantly. For a
ballistic 1d ring of radius $r_0$ with azimuthal field texture  the latter
time corresponds to the period of the orbital motion along a round path
$\Gamma$, and the condition for adiabaticity reads
\be
\label{1Dadiab}
\fr{\omega}{\omega_{\rm s}}\ll 1 \; .
\ee
Here, $\omega=v_{\rm F}/r_0$ is the orbital frequency of an electron
with Fermi velocity $v_{\rm F}$.  Eq. (\ref{1Dadiab}) is deduced in the
1d case \cite{stern}
by comparing the off-diagonal matrix elements 
$\la \psi^{\uarr (\darr)}|H^{\uarr \darr (\darr \uarr)}|\psi^{\darr (\uarr)}\ra$ 
of the corresponding 1d Hamiltonian, i.e.\ the terms proportional to $\sin \alp$ 
in Eq. (\ref{qhamndB-sym}) for fixed $r$, with the diagonal  
matrix elements $\la \psi^{\uarr (\darr)}|H^{\uarr
(\darr)}|\psi^{\uarr (\darr)} \ra$.
Following the same procedure we obtain conditions for adiabaticity in the 
rotationally symmetric 
2d case by evaluating the matrix elements of the Hamiltonians
(\ref{qhamdB}) and (\ref{qhamndB-sym}). The term proportional to
$\sin \alp$ in Eq. (\ref{qhamndB-sym}) gives rise to a
condition equivalent to Eq. (\ref{1Dadiab}) by replacing $r_0$ and $B$ 
by the respective mean values of radius and magnetic field of the 2d ring. 

The additional requirement that the second term on the rhs of 
Eq.~(\ref{qhamndB-sym}) containing derivatives of the  angle $\alp$ is small, 
leads to a complementary adiabaticity condition.
Evaluating the respective matrix elements in Eq.~(\ref{qhamndB-sym}) yields
\be
\label{nd-2d-1d}
\pm \fr {\hbar^2}{4m*} \int \fr {\partial \alp}{\partial r}  \li( \psi^{\uarr
(\darr)*} \fr {\partial \psi^{\darr (\uarr)}} {\partial r}- \fr {\partial
\psi^{\uarr (\darr)*}} {\partial r}  \psi^{\darr (\uarr)}\re) r {\rm d}r 
{\rm d} \varphi \; .
\ee 
To obtain analytical estimates for this contribution we assume in the following
that the system has the form of an annulus (Fig.\ \ref{fig:model}(a)) and
that the angular momentum is conserved (assuming a weak coupling with the leads). 
Then the  orbital and radial motion are separable, and we can write the spatial
wave functions as products 
$\psi^\uarr_{nl}(\vec{r})= \exp(il\varphi) \phi^\uarr_{nl}(r)$ and 
$\psi^\darr_{n'l'}(\vec{r})= \exp(il'\varphi)
\phi^\darr_{n'l'}(r)$. Here $\phi^\uarr_{nl}$ ($\phi^\darr_{n'l'}$) are the 
solutions of the corresponding radial
Schr\"odinger equation for spin up (down) with transverse mode  quantum numbers 
$n$ ($n'$) and azimuthal quantum numbers $l$ ($l'$), respectively. 
The matrix elements  (\ref{nd-2d-1d}) are non-zero only for $l=l'$.
Adiabaticity requires the absolute value of these terms to be
small compared to the Zeeman energy $\mu B$. To evaluate this
condition we further consider states $\phi_{nl}$ in a 2d ring of width $d=R_2-R_1$ and 
mean radius $r_0=(R_1+R_2)/2$ such that its aspect ratio $d/r_0 \ll \pi n/l$. This
allows us to approximate the radial eigenstates (for hard-wall boundary
conditions) as $(1/\sqrt{\pi d \ r}) \sin (k_n r)$, independent of $l$,
with $k_n=\pi n/d$.  With these approximations the resulting adiabaticity 
condition reads
\be
\label{nd-2d-1d-sym}
\left|
\frac{ \omega_n}{ \omega_{\rm s}} I(\alp; n, n') 
-\frac{ \omega_{n'}}{ \omega_{\rm s}} I(\alp; n', n) \right| \ll 1 \; ,
\ee
% \Bigg{|}\int_{\rm R_1}^{\rm R_2} \fr {\partial \alp}{\partial r}
%  \ \Big{[} \omega_n \ \sin[k_{n'}(r-R_1)] \ \cos[k_n(r-R_1)] - \omega_{n'} \
% \sin[k_n(r-R_1)] \ \cos[k_{n'}(r-R_1)] \Big{]} \ dr \Bigg{|} \ll \omega_{\rm s}
with
\be
\label{integral}
 I(\alp; n, n') = 
\int_{\rm R_1}^{\rm R_2} \fr {\partial \alp}{\partial r} \
\cos[k_n(r-R_1)] \ \sin[k_{n'}(r-R_1)] 
 \ {\rm d} r
\ee
and $\omega_n=\hbar k_n/(2m^* d)$ being the bounce frequency
in radial direction associated with the mode $n$.

The adiabaticity condition (\ref{nd-2d-1d-sym}) with (\ref{integral}) 
depends on the specific
form of $\alp(r)$. To obtain a simple estimate we assume a monotonic behavior for $\alp(r)$
and choose for simplicity  $\alp= \alp_0 \
\exp(-\delta \ r)$ with $\delta>0$\cite{note5}. This gives as an upper bound
for any pair $n,\ n'$ 
\be
\label{2Dadiab-a}
{
\fr{\omega_\alp}{\omega_{\rm s}} \equiv 
\fr{\omega_{\rm N}}{\omega_{\rm s}} \ |\Delta \alp|  \ll 1 \; ,
}
\ee
where $\omega_{\rm N}=v_{\rm F}/d$ is the bounce frequency for the
highest mode $N={\rm Int}[k_{\rm F} d/\pi]$, $\Delta \alp = \alp(R_2) -  \alp(R_1)$,
and we have defined a mean angular frequency 
$\omega_\alp=|\Delta \alp|\omega_N$.
The condition (\ref{2Dadiab-a}) shows in a simple manner that the
radial motion is also subject to constraints in order to satisfy
adiabaticity. For a given Fermi energy both the field
texture and field strength are generally relevant.

In the numerical applications below the condition (\ref{2Dadiab-a}) is
easily satisfied, while the condition (\ref{1Dadiab}) remains as the
stronger constraint. We note that the adiabaticity
requirements must be satisfied for every value of the uniform field
$B_0$. Hence, we use $B_{\rm i}$ instead of $B$ for checking
Eq. (\ref{1Dadiab}).  

To see whether adiabaticity is achieved in ballistic devices built
from high-mobility heterostructures we evaluate Eq.\ (\ref{1Dadiab}) 
for typical samples of ring 
geometry\cite{weiss,lindelof} with $r_0\approx300$ nm and a width of the rings
corresponding to five open transverse modes.
For instance for InAs samples ($g^*\approx 15$ and $m^*/m_0\approx0.023$) 
adiabaticity requires a magnitude of at least $B_{\rm i}= 1$~Tesla.
Nonuniform fields $\vec{B}_{\rm i}$ varying on mesoscopic scales
have been recently achieved by placing a micromagnet at the center of ballistic
rings. The micromagnet creates a tilted, rotationally symmetric field
at the level of the 2DEG\cite{weiss}. 
Whereas the above estimated magnitude seems rather large for 
mesoscopic sources of magnetic fields, it has been recently reported
that using such micromagnets one can indeed achieve
magnetic inhomogenities up to 1 Tesla\cite{geim} which open
up the possibility to measure Berry phases.

%
%%%%%%%%%%%%%%%%%%%%%%%%%%%%%%%%%%%%%%%%%%%%%%%%%%%%%%%%%%%%%%%%%%%%
%%%%%%                  NEW SECTION
%%%%%%%%%%%%%%%%%%%%%%%%%%%%%%%%%%%%%%%%%%%%%%%%%%%%%%%%%%%%%%%%%%%%
%

\section{Model and Quantum Transport Calculations}
\label{sec:approach}

\begin{figure}
\begin{center}
\leavevmode
\epsfxsize=0.35\textwidth
\epsfbox{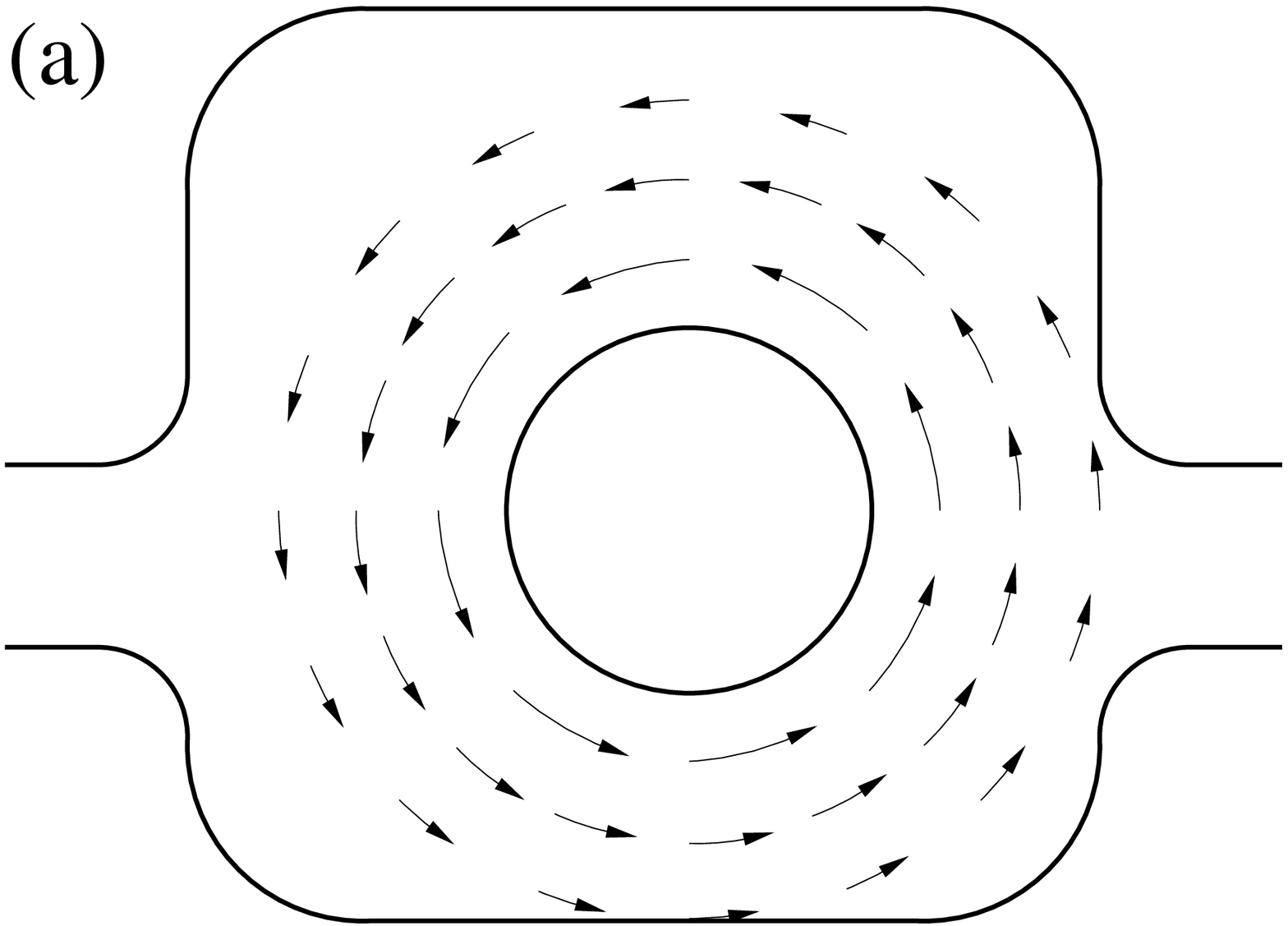}
\hspace*{2.5cm} 
% \vspace*{1cm}
\leavevmode
\epsfxsize=0.27\textwidth
\epsfbox{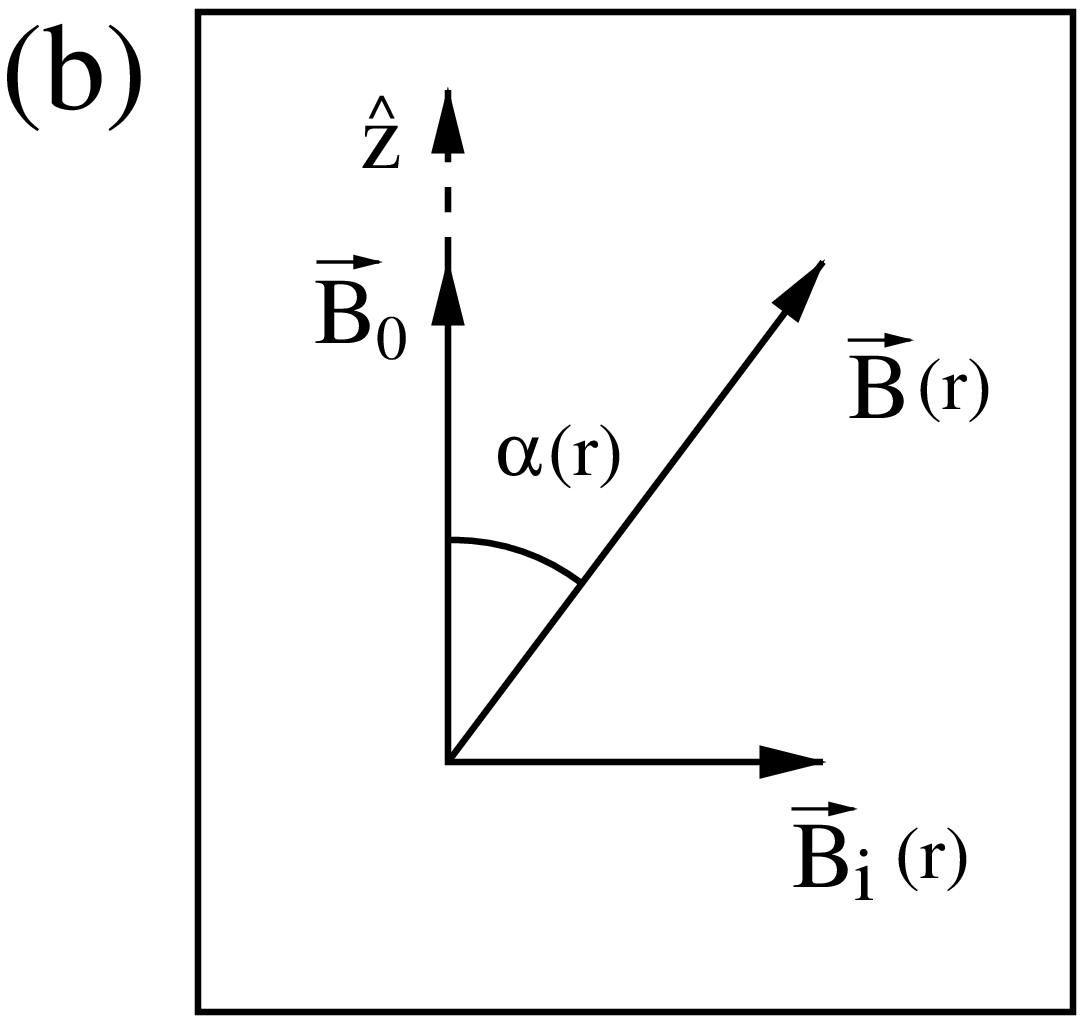}
\end{center}
\caption{
(a) 
Sketch of an inhomogenous magnetic field, Eq.\ (\ref{Bs}), giving rise to
Berry phase effects in quantum transport;
(b) the total magnetic field $\vec{B}(r)$ is composed of an
inhomogenous field $\vec{B}_{\rm i}=B_{\rm i}(r) \ \hat{\varphi}$ as in 
(a) and a perpendicular homogenous field $\vec{B}_0$.
}
\label{fig:Bfield}
\end{figure}

We study numerically quantum transport through 2d rings coupled to two leads
and, more generally, doubly-connected structures of the type shown 
in Fig.~\ref{fig:model}(b).
As a model of the nonuniform magnetic field $\vec{B}_{\rm i}$ we use a circular
field as depicted in Fig.~\ref{fig:Bfield}(a). Such a field configuration
can be viewed as being 
generated by an electrical current in $\hat{z}$-direction or it can 
be achieved in ferromagnetic rings\cite{experiment}. 
For our numerical calculations we choose as symmetry axis the $\hat{z}$-axis
through the center of the inner disk of the microstructure and use
\be
\label{Bs}
\vec{B}_{\rm i}(\vec{r})= B_{\rm i}(r) 
\ \hat{\varphi}= \fr{a}{r}  \ \hat{\varphi} \; .
% \fr{1}{r} \ \fr{\mu^*}{2\pi} \ I \ \hat{\varp hi}
\ee
(For the model of a current $I$ generating the inhomogenous 
field, $a=\mu^\ast I/2\pi$.)
The electromagnetic vector potential corresponding to
$\vec{B}_{\rm i}$ as chosen in Eq.~(\ref{Bs}) does not contribute to
$\phi_{\rm eff}^{\uarr (\darr)}$ in Eq.~(\ref{phi-eff}), i.e. $\phi_{\rm i}=0$. 
However, $\vec{B}_{\rm i}$ gives rise to a geometrical phase in the same manner as
the field of a micromagnet. 
The angle $\alp(r)$ entering into the expression (\ref{Ag}) for
the geometric vector potential is the tilt 
angle of the total field arising from $\vec{B}_{\rm i}$ and an additional
perpendicular homogenous field $\vec{B}_0$, Fig.~\ref{fig:Bfield}(b).

We compute the zero-temperature conductance $G$ of the microstructures
in the linear-response regime within the Landauer framework\cite{landauer}
which states that $G$ is proportional to the transmission $T$ through the
system.
For a microstructure with two leads of width $W$ attached supporting each 
$N={\rm Int}[k_{\rm F}W/\pi]$ transverse modes the conductance 
for spin-independent quantum transport reads
\be
\label{landa-ns}
G(E_{\rm F}, \vec{B}) = {\rm g}_{\rm s} \fr{e^2}{h} T(E_{\rm F},\vec{B}) = 
{\rm g}_{\rm s} \frac{e^2}{h} \sum_{n,m=1}^N |t_{nm}|^2  \; .
\ee
The $t_{nm}$ denote transmission amplitudes between 
incoming ($m$) and outgoing ($n$) channels in the leads.
They are obtained by projecting the Green function ${\cal G}$ of the system
onto the transverse mode functions $\phi_{m}(y)$ and $\phi_{n}(y')$ 
in the leads \cite{fisher}:
\be
\label{t-nm}
{
t_{nm} = -i\hbar(v_n v_m)^{1/2}
\int dy' \int dy\,\phi_n^\ast(y') \phi_m(y) \ 
{\cal G}(x',y';x,y;E_{\rm F};\vec{B})\; .
}
\ee
Here, the $y$- and $y'$-integrations are performed along transverse cross sections 
of the left and right lead located at (horizontal) positions $x$ and $x'$ 
in the lead.
$v_n$ is the longitudinal velocity of propagation of an asymptotic channel 
wave function with transverse mode $n$.

In Eq.~(\ref{landa-ns}),
the prefactor ${\rm g}_{\rm s} = 2$  takes into account the 
spin degrees of freedom in the case of spin-independent transport. In the
presence of spin coupling to a magnetic field the generalized expression for
the conductance reads
\be
\label{landa-s}
G = \frac{e^2}{h} \sum_{n,m=1}^N \li(\dpl|t_{nm}^{\uarr\uarr}|^2+
 |t_{nm}^{\uarr\darr}|^2+|t_{nm}^{\darr\darr}|^2+|t_{nm}^{\darr\uarr}|^2\re) \; .
\ee
Working within the adiabatic approximation we {\it neglect} the offdiagonal
terms in Eq.\ (\ref{landa-s}), computing only the amplitudes
$t_{nm}^{\uarr\uarr}$ and $t_{nm}^{\darr\darr}$,
with well defined spin polarization within the cavity.
These terms are calculated independently, according to the decoupling of the
two corresponding Hamiltonians $H^{\uarr(\darr)}$ in Eq.\ (\ref{qhamdB}).
Hence, in the adiabatic limit unitarity imposes for each spin direction
$\sum_{n,m}^N(|t_{nm}^{\uarr\uarr(\darr\darr)}|^2+
|r_{nm}^{\uarr\uarr(\darr\darr)}|^2)=T^{\uarr(\darr)}+R^{\uarr(\darr)}\equiv N$,
where $r_{nm}$ are the corresponding reflection amplitudes 
(for their precise definition see e.g.\ Ref.~\cite{baranger1}).
 
To compute the conductance  
we first calculate the Green functions ${\cal G}^{\uarr(\darr)}$
for the Hamiltonians  $H^{\uarr(\darr)}$
numerically on a grid within a tight-binding model using a recursive
method\cite{ferry} and then perform the integrals (\ref{t-nm}).
The different cavity geometries considered are introduced via
the potential $V$ in Eq.\ (\ref{qhamdB}) and implemented by using
hard-wall boundary conditions.

The effective potential $V_{\rm eff}$, Eq.~(\ref{Veff}), which 
enters into  $H^{\uarr(\darr)}$ contains an $\alp$-dependent geometrical
term and the Zeeman term. The geometrical part is usually  small
compared to the Fermi energy as we see if we express it in energy scaled 
units (after dividing by $\pi$ times the mean level spacing 
$\Delta = \hbar^2/2\pi m^* r_0^2$):
\be
\label{ineq}
\fr{1}{4}\li[\sin^2\alp + r_0^2 \ \left(\fr{\partial 
\alp}{\partial r}\right)^2  \re] \ll
(k_{\rm F} \ r_0)^2 \; .
\ee
This relation is usually justified since $k_{\rm F}  r_0 = 2\pi r_0/\lambda_{\rm F}$
is large in the mesoscopic regime and the left hand side of Eq.~(\ref{ineq}) 
is typically of order one (if $\alp$ does not vary too fast with $r$). Hence we can 
neglect this term in our calculations.

Likewise, for the scaled Zeeman energy $\mu B$  we find
\be 
k_{\rm F} \ r_0 \ll g^*  \fr{m^*}{m_0}  \fr{\pi r_0^2 B}{\phi_0}
\ll (k_{\rm F} \ r_0)^2 \; ,
\ee 
where the first inequality represents the scaled adiabaticity condition (\ref{1Dadiab}).
Magneto resistance experiments on InAs devices
have shown that the Zeeman spin splitting is not 
manifested up to a field strength of about 1.5~Tesla what is compatible with 
our approximations \cite{ensslin}. 
Therefore, for the systems and quantities studied in this paper we can
neglect also the Zeeman term in our numerical calculations.
In particular in our study of the energy-averaged magneto resistance
the Zeeman splitting does not play a role.
We note, however, that conductance fluctuations in individual
ballistic systems
can generally be sensitive to energy variations on scales of $\mu B$. 

%%%%%%%%%%%%%%%%%%%%%%%%%%%%%%%%%%%%%%%%%%%%%%%%%%%%%%%%%%%%%%%%%%%%%

\section{Numerical Results and Discussion}
\label{sec:results}

To achieve a better understanding of Berry phase effects in the ballistic 
mesoscopic regime we consider different representative 2d
cavity geometries and address quantum phenomena for transport through
single systems as well as ensemble averages. Thereby we
study signatures of geometrical phases in Aharonov-Bohm oscillations as well as 
in weak-localization phenomena.
 
The results are organized as follows: In Sec. \ref{results-se}, 
we first present numerical calculations of the  magneto conductance
through ring geometries at \emph{fixed} Fermi energy. Two complementary cases are
analyzed: 
(i) a rotationally symmetric ring, Fig.~\ref{fig:model}(a), 
as an example for a quasi-1d configuration
(one open channel, small aspect ratio $d/r_0$); 
(ii) an asymmetric ring-type geometry as shown in 
Fig.~\ref{fig:model}(b) with a mean aspect ratio corresponding to
several open transverse channels.
In Sec.\ \ref{results-ea} we then summarize our results for the
\emph{energy-averaged} magneto resistance representing 
an ensemble average for microstructures varying in size. 
% Results are again given for the geometry in Fig.\ref{fig:model}(b). 

\subsection{Magneto conductance for single systems}
\label{results-se}

\begin{figure}
\begin{center}
\leavevmode
\epsfxsize=0.6\textwidth
\epsfbox{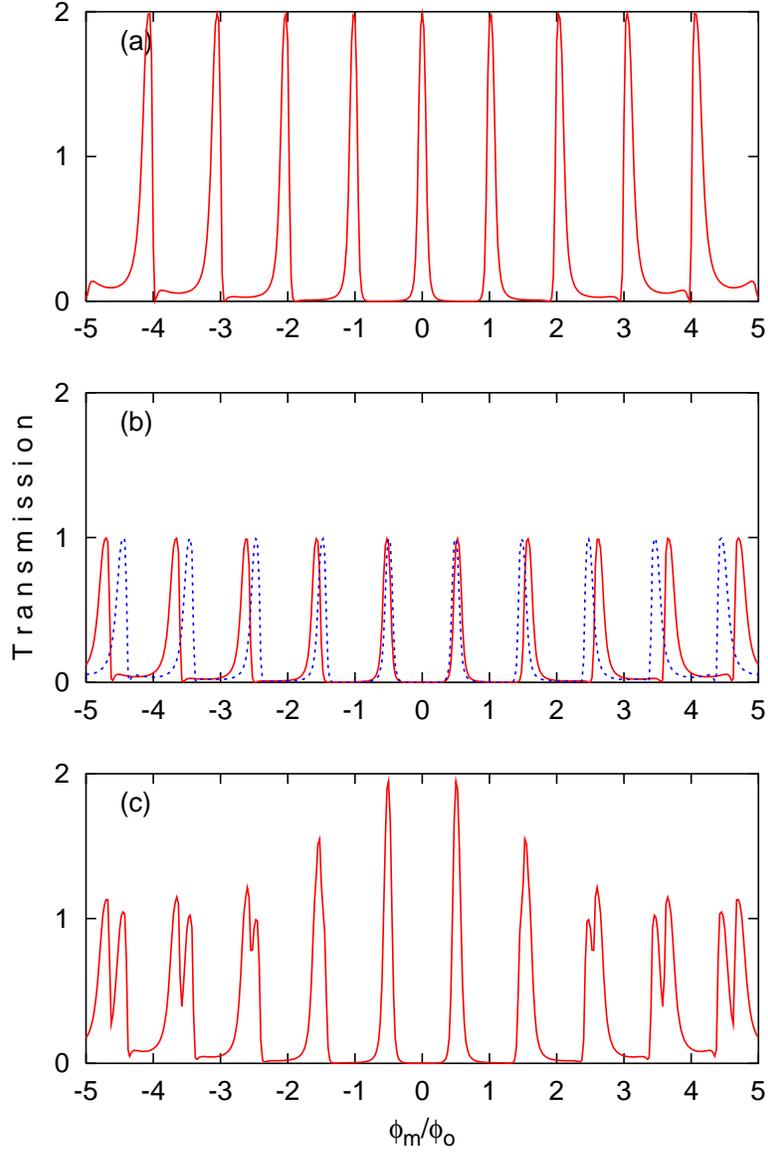}
\end{center}
\caption{
Quantum transmission as a function of mean flux 
$\protect \phi_m= \pi r_0^2 B_0$ through a ballistic 
ring (Fig.\ \ref{fig:model}(a), small aspect ratio) 
for a Fermi energy corresponding to one open
transverse mode in the ring and in the leads. 
(a) Aharonov-Bohm oscillations for the
case without inhomogenous field (the spin degree of freedom is included by
means of a factor 
 $\protect {\rm g}_{\rm s}= 2$);
(b) effect of the geometrical phase owing to the
spin coupling to an inhomogenous magnetic field as
sketched in Fig.\ \ref{fig:Bfield}(b). The solid
(dashed) curve shows the transmission coefficient
$T^\uarr$ ($T^\darr$) of spin-up (spin-down) electrons;
(c) total transmission $T^\uarr + T^\darr$.
}
\label{fig:3}
\end{figure}

\begin{figure}
\begin{center}
\leavevmode
\epsfxsize=0.6\textwidth
\epsfbox{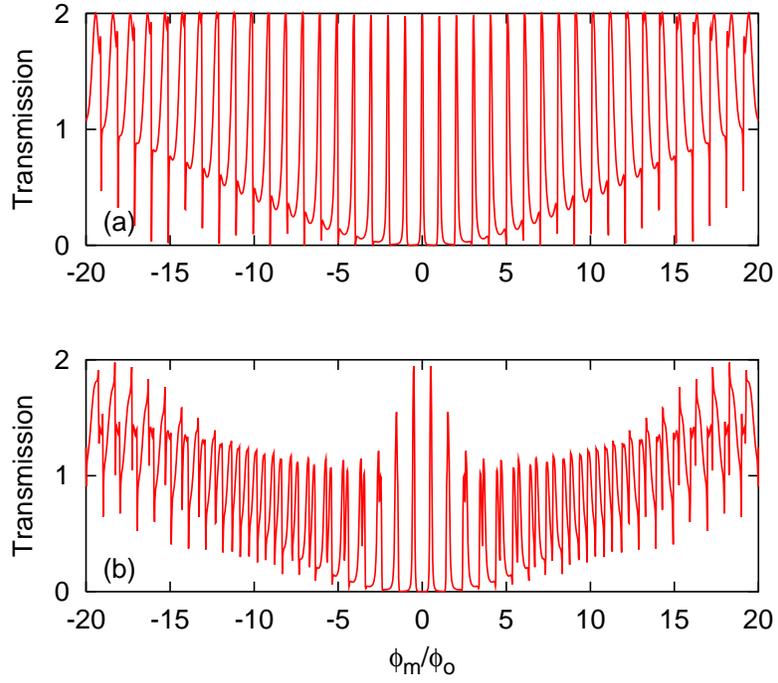}
\end{center}
\caption{
Transmission for the same ring geometry and parameters
as in Fig.\ \ref{fig:3} for a wider range of the mean flux
without  (a) and  with (b) inhomogenous field.
}
\label{fig:4}
\end{figure}

Figs.\ \ref{fig:3} and \ref{fig:4} show our results for the quantum
transmission $T$ as a function of the mean flux $\phi_{\rm m} = \pi r_0^2 B_0$ 
through an Aharonov-Bohm ballistic ring as the one shown in Fig.\ \ref{fig:model}(a).
The geometry parameters used are the mean radius $r_0/W=(R_1+R_2)/(2W)=2.7$
and the aspect ratio $d/r_0 = (R_2 - R_1)/r_0 = 0.22$. The dimensionless Fermi wave number is 
$k_{\rm F}W/\pi=1.8$, hence the
leads and the ring support a single open channel
($N={\rm Int}[k_{\rm F}W/\pi]=1$). This situation 
is close to the case of a  1d ring
\cite{loss,stern,pichugin}. Fig. \ref{fig:3}(a) shows the transmission for 
 $B_{\rm i}=0$, i.e. only  the external homogenous field is present. 
As expected one observes regular Aharonov-Bohm
oscillations with a well defined period of one flux quantum, characteristic for
1d rings. However, a smooth modulation of the amplitude arises
when studying a wide range in $B_0$, as shown in Fig.\ \ref{fig:4}(a). 
This results from the finite width of the ring. 
The spin degrees of freedom are taken into account by the factor ${\rm g}_{\rm s} = 2$
in Eq.\ (\ref{landa-ns}). 

Fig.~\ref{fig:3}(b) depicts the effect of the inhomogenous field
$B_{\rm i}$, Eq.~(\ref{Bs}), on the conductance. 
The numerically calculated spin-dependent transmission 
coefficients $T^\uarr$ and $T^\darr$ are shown as the solid and 
dashed curve. Owing to the effect of the Berry phase, two new features
arise in the transmission profiles:
 (i) a phase shift of $\phi_0/2$ at $\phi_{\rm m}=0$ for
both $T^\uarr$ and $T^\darr$ with respect to the case $B_{\rm i}=0$,
Fig. \ref{fig:3}(a). 
(ii) the periods  $\phi^{\uarr(\darr)}$ of the oscillations are modified 
with respect to the AB period $\phi_0$ 
in such a way that $\phi^\darr \le \phi_0 \le \phi^\uarr$. 
This behavior, which was predicted for 1d rings\cite{stern},
shows up in the  total transmission, $T^\uarr + T^\darr$, 
as splitting of the peaks in Fig.\ \ref{fig:3}(c)
and leads to a pronounced modulation of the oscillation amplitude on
larger scales of $\phi_{\rm m}$, Fig.\ \ref{fig:4}(b). 

(i) The phase shift at zero flux is related to the fact that
one has $\cos \alp =0$ for vanishing external field (Fig.~\ref{fig:Bfield}).
Thus the Berry phase, Eq.~(\ref{Bphase}), is $\gamma^{\uarr (\darr)} = \pi$
for $B_0 = 0$, and its contribution to the effective flux $\phi_{\rm
eff}$ in Eq. (\ref{phi-eff}) is $\phi_0\gamma^{\uarr(\darr)}/2\pi=\phi_0/2$.
This holds for both spin polarizations and for any path $\Gamma$ around the ring.
Hence, owing to the geometrical phase, 
$T^{\uarr(\darr)}$ exhibits a minimum at $\phi_{\rm m}=0$ 
instead of the peak for the case with $B_{\rm i}= 0$.
We note, however, that even in the pure AB case
the magneto conductance peak positions and peak profiles depend also
on the Fermi energy\cite{lindelof,imry}. Hence, including the different
shifts in energy arising from $V_{\rm eff}^\uarr$ and $V_{\rm eff}^\darr$
in Eq.~(\ref{Veff}) will presumably render the effect (i) less clear.
This problem disappears in the case of the averaged magneto conductance. 

(ii) The modified period of the spin-dependent conductance
 oscillations is a consequence of the
dependence of the geometrical phase  $\gamma^{\uarr(\darr)}$ on
$B_0$ through $\cos \alp$ (Eq.~(\ref{Bphase})). 
According to Eq.~(\ref{phi-eff})
the effective phase accumulated by an electron along a closed
path $\Gamma$ is
% , owing to the effective vector potential,
$\varphi_{\rm eff}^{\uarr(\darr)}=2\pi
\phi_{\rm eff}^{\uarr(\darr)}/\phi_0$. Upon varying
  $B_0$, this phase changes with a spin-dependent rate
\be
\omega_\Gamma^{\uarr(\darr)}=\fr{\partial \varphi_{\rm eff}^{\uarr(\darr)}}{\partial
B_0}= \fr{2\pi}{\phi_0} \ {\cal A}_{\Gamma}\mp \oint_{\Gamma} \fr{1}{2r}
\ \fr{\partial \cos \alp(r,B_0)}{\partial B_0} \ \hat{\varphi} \cdot \vec{{\rm d}l}
\ee
which defines the frequency of the magneto conductance oscillations. The first
term corresponds to the electromagnetic flux while the second one is of
geometrical origin. In the case of a quasi-1d ring of mean radius
$r_0$ the frequency of the oscillations can be approximated as 
\be
{
\omega^{\uarr(\darr)} \simeq  \omega_0 \mp \pi \
\fr{\partial \cos \alp(r_0, B_0)}{\partial B_0}
}
\ee
where $\omega_0= 2\pi (\pi r_0^2)/\phi_0$ corresponds to the case $B_{\rm i}=0$. 
This splitting in frequency 
is found in the numerical results shown in Fig.~\ref{fig:3}(b,c).
Taking into account that $\partial \cos \alp/\partial B_0 \ge 0$ we finally find 
$ \omega^{\uarr} \le \omega_0 \le \omega^{\darr} $.
For fixed $B_{\rm i}$ one has $\omega^{\uarr(\darr)} \rightarrow  \omega_0$
for $B_{\rm i}/B_0 \rightarrow 0$ giving rise to a compressed or spread
set of conductance oscillations. The splitting of the frequencies is visible
for a broader flux range in Fig.~\ref{fig:4}(b) as a dephasing between
the different spin polarizations and a modulation of the
amplitude of the total transmission, $T^\uarr+T^\darr$. The flux
dependence of the amplitude modulation can be qualitatively explained
within a sinusoidal model employing\cite{note3}
\bea
T^{\uarr}+T^{\darr} \sim \cos \varphi_{\rm eff}^{\uarr} + \cos
\varphi_{\rm eff}^{\darr}= 
-2 \ \cos (2\pi \phi/\phi_0) \cos [\pi \cos \alp(B_0)] \; .
\eea
Nodes in the amplitude correspond to field strengths where $\cos \alp =\pm 1/2$. 
The distance between the two existing nodes is $(2/\sqrt{3}) B_{\rm i}$
increasing linearly with the inhomogenous field.

% The last curiously indicates that a very large
% $B_{\rm i}$ (what improves adiabaticity) would complicates the detection of
% the Berry phase by the retiring of the \emph{nodes} and the similitude between
% $\omega^{\uarr}$ and $\omega^{\darr}$ ($\partial \cos \alp/\partial
% B_0 \rightarrow 0$ when $B_{\rm i} \rightarrow \infty$). Then, a huge
% range should  be explore in $B_0$ for having a possitive result. This would
% make not possible the splitting of the extremes due to the slow rate of
% accumulation of geometric phase. 

In order to generalize the results to more realistic systems 
we also performed calculations on asymmetric 2d ring structures which
support several open channels. The
numerical conductance calculations were performed for the 
ring-type geometry shown in Fig.~\ref{fig:model}(b).
The asymmetry is introduded by means of a displacement of the inner object 
from the center and by a shift of the leads.
The geometry parameters in the calculations are $L_{\rm x}/W=L_{\rm y}/W=3.8$,
the mean radius $r_0/W=(R_1+R_2)/(2W)=1.7$, and the mean aspect ratio
$d/r_0 = (R_2-R_1)/r_0 = 0.35$, with $R_2 \equiv \sqrt{L_{\rm x}L_{\rm
y}/\pi}$. The dimensionless wave number is $k_{\rm F}W/\pi=4.85$, corresponding
to four contributing channels in the leads, 
while the number of open modes is not well defined in the ring, 
varying between two and three effective channels. 

\begin{figure}
\begin{center}
\leavevmode
\epsfxsize=0.6\textwidth
\epsfbox{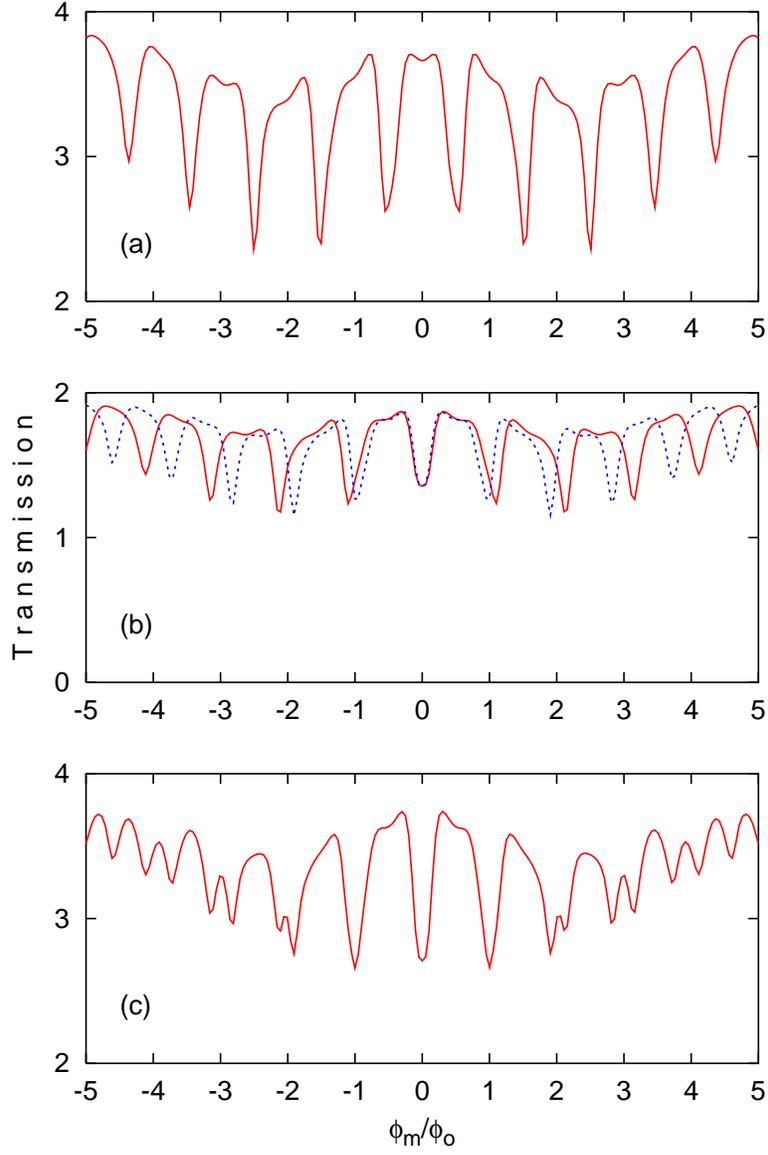}
\end{center}
\caption{
Quantum transmission as a function of mean flux $\phi_m=
\pi r_0^2 B_0$ through an asymmetric 
ring geometry as shown in Fig.\ \ref{fig:model}(b)
for a Fermi energy corresponding to four open
modes in the leads.
(a) Aharonov-Bohm type oscillations for the
case without inhomogenous field (the spin degree of freedom is included by
means of a  factor $\protect {\rm g}_{\rm s} = 2$);
(b)
effect of the geometrical phase for finite $B_{\rm i}$:
The solid (dashed) curve shows the 
transmission coefficients
$T^\uarr$ ($T^\darr$) of spin-up (spin-down) electrons;
(c) total transmission $T^\uarr + T^\darr$.
}
\label{fig:5}
\end{figure}

In Fig.~\ref{fig:5} we present the results for the quantum transmission
displayed in the same way as in Fig.\ \ref{fig:3} for the quasi-1d case. 
For $B_{\rm i} = 0$,  Fig. \ref{fig:5}(a), we observe slightly irregular
oscillations, reduced in amplitude with respect to the quasi-1d case,
owing to the asymmetry of the structure  and interference 
between the various contributing channels. The particular shape of the transmission
profile is strongly energy-dependent.  

Nevertheless, for finite $B_{\rm i}$ one finds again a splitting
of the period of the oscillations into spin-polarized contributions, as shown
in Fig.~\ref{fig:5}(b), i.e. $\phi^{\darr} \le \phi_0 \le \phi^{\uarr}$.  
As in the 1d case, a phase shift  by $\phi_0/2$ is visible in the transmission 
oscillations at $\phi_{\rm m}=0$ in Fig.~\ref{fig:5}(b),
which is  again a signature of the geometrical phase, as discussed above.

In Fig.~\ref{fig:5}(c) we see that the splitting in period is still
observable in the total transmission,
$T^{\uarr}+T^{\darr}$, similar to the quasi-1d case.
However, a study of a wider range in $B_0$, not presented here,
shows that an amplitude modulation generated
by the Berry phase, corresponding to Fig.~\ref{fig:4}(b),
is hardly distinguishable from a modulation which
is already present in the AB background due to the finite width of the
ring or larger number of open channels, respectively.

Finally, a comment on the observed symmetry property of the transmission 
with respect to inversion of the uniform field, 
i.e.\ $T^{\uarr(\darr)}(B_0)=T^{\uarr(\darr)}(-B_0)$, is due. The effective
Hamiltonians (\ref{qhamdB}) for spin-polarized electrons describe a
 system which is subject to the action of a spin-dependent,
(inhomogenous) effective magnetic field $\vec{B}_{\rm eff}^{\uarr(\darr)}=\vec{\nabla} \times
\vec{A}_{\rm eff}^{\uarr(\darr)}$ (see Eq.~(\ref{Aeff})). For electron 
motion in the x-y plane, the reciprocity relations\cite{buttiker,baranger2} 
for two-terminal transport in the linear regime state that in this situation  
$T^{\uarr(\darr)}(B_{\rm eff}^{z \ 
\uarr(\darr)})=T^{\uarr(\darr)}(-B_{\rm eff}^{z \ \uarr(\darr)})$,
where $B_{\rm eff}^{z \ \uarr(\darr)}$ is the z-component of the effective
field $\vec{B}_{\rm eff}^{\uarr(\darr)}$. However, it can be
proved that
$B_{\rm eff}^{z \ \uarr(\darr)}(B_0)=-B_{\rm eff}^{z \
\uarr(\darr)}(-B_0)$ in the case of an  inhomogenous field
$\vec{B}_{\rm i}$ as defined in Eq.~(\ref{Bs})\cite{note4}.

%%%%%%%%%%%%%%%%%%%%%%%%%%%%%%%%%%%%%%%%%%%%%%%%%%%%%%%%%%%%%%%%%%%%%

\subsection{Energy-averaged magneto conductance}
\label{results-ea}

As mentioned above the magneto conductance profiles for transport
through single systems are energy dependent. In this
section we therefore address the question whether the more robust
energy-averaged conductance exhibits features of the geometrical phase.
A consideration of the averaged conductance has also the 
advantage that additional fluctuations in the conductance
which may arise from the spin-dependent background 
$V_{\rm eff}^\uarr$ and $V_{\rm eff}^\darr$, Eq.~(\ref{Veff}), 
are washed out.
An energy average is experimentally realized by an ensemble average
over microstructures of similar shape but varying size.

\begin{figure}
\begin{center}
\leavevmode
\epsfxsize=0.6\textwidth
\epsfbox{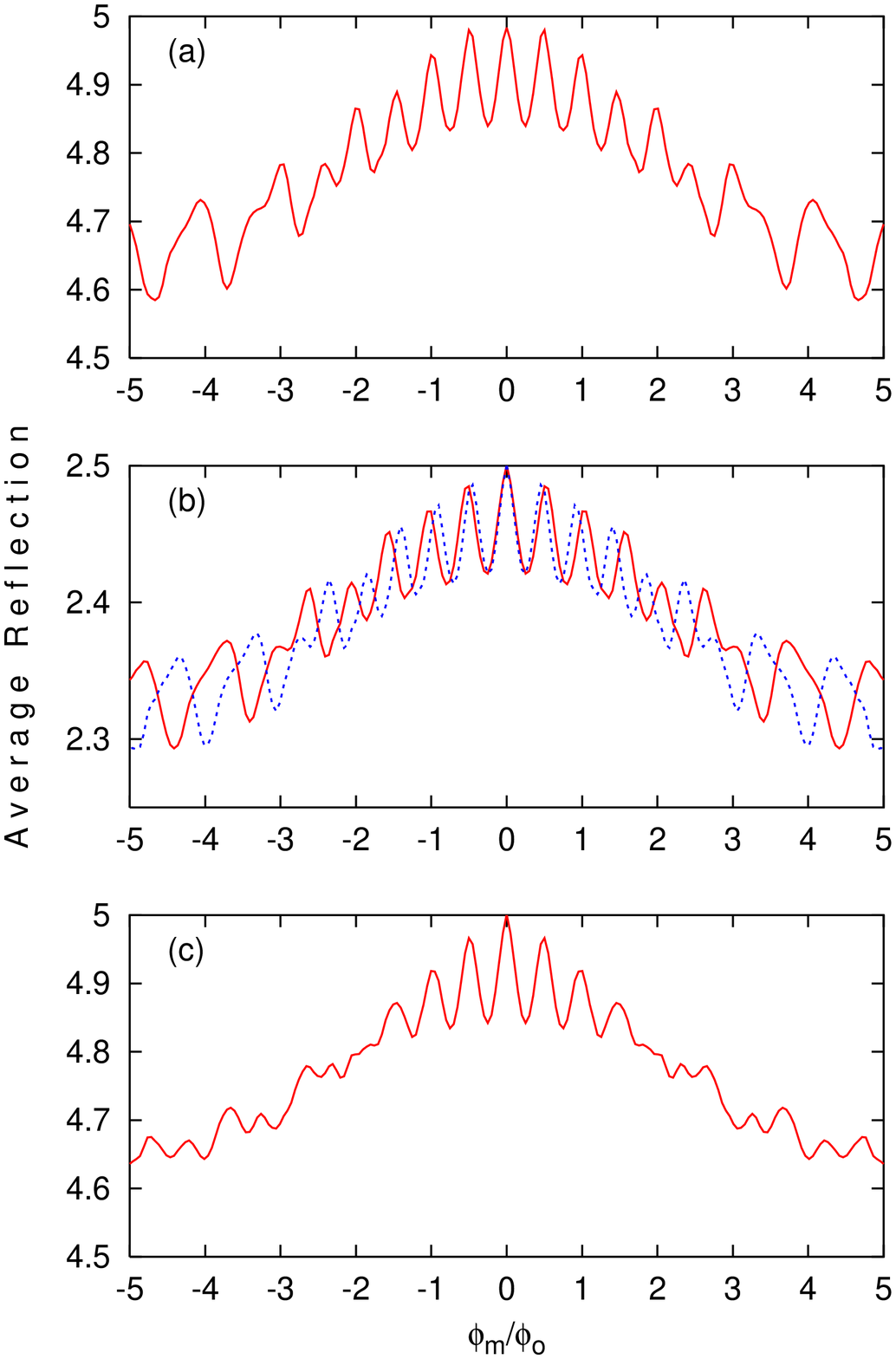}
\end{center}
\caption{Effect of the Berry phase on weak localization 
in ballistic rings. The energy-averaged reflection coefficient 
is shown for the same geometry as in Fig.~\ref{fig:5}.
(a) $\phi_0/2$-oscillations for  $B_{\rm i}=0$
(the spin degree of freedom is included by means of a 
factor ${\rm g}_{\rm s} = 2$);
(b) effect of the geometrical phase for finite $B_{\rm i}$:
the solid (dashed) curve shows the averaged reflection
$\langle R^\uarr \rangle $ ($\langle R^\darr \rangle$);
(c) total averaged reflection $\langle R^\uarr \rangle+\langle R^\darr \rangle$.
}
\label{fig:6}
\end{figure}

Here we present results of numerical calculations on the
energy-averaged magneto conductance for an ensemble of
asymmetric ring structures  as in Fig.~\ref{fig:model}(b) for
the same geometry parameters as used in Fig.~\ref{fig:4}. 
The average is performed over 50 different energy values
in a window $\Delta E$ corresponding to the whole
fourth conducting channel, i.e. $N={\rm Int}[k_{\rm F}W/\pi]=4$ within
$\Delta E$.  It is convenient to express the results in terms of an 
average reflection (AR) to discuss weak-localization phenomena. Our
results are summarized in  Fig.~\ref{fig:6}. Panel (a)
shows the AR for $B_{\rm i}=0$ where no geometrical phase effects
are present. We find oscillations on top of a broad 
peak centered at $B_0 = 0$. 
A sequence of oscillations close to the maximum exhibits
a period of $\phi_0/2$ which turns into oscillations with 
period $\phi_0$ at larger external field.
As a consequence of the averaging the amplitude of the oscillations is
reduced by a factor of about 5 with respect to the  single-energy case,
Fig.~\ref{fig:5}(a). 

The maximum in AR at $B_0=0$ and the sequence of oscillations  
with period $\phi_0/2$ can be ascribed to weak-localization phenomena.
The $\phi_0/2$ oscillations which  we find here for ballistic rings 
are analogous to the Altshuler-Aronov-Spivak oscillations in
disordered conductors\cite{spivak}. In a semiclassical 
picture\cite{baranger1,reviews}
these oscillations can be qualitatively understood and have
been associated with pairs of time reversed, backscattered paths
enclosing the inner disk\cite{tanaka}.
For fluxes larger than $|\phi_{m}| \approx 2.5$ the
time reversal symmetry is broken, giving rise to the periodicity of $\phi_0$
as for the usual AB effect.
The same crossover from $\phi_0/2$ to $\phi_0$ periodicity has
been recently experimentally observed for ensembles of 2d ballistic 
rings built from semiconductor heterostructures\cite{weiss2}. 

In Figs.~\ref{fig:6}(b,c)  we present our numerical
results for a finite $B_{\rm i}$, where effects of Berry phases are expected.
 The solid and dashed line in Fig.~\ref{fig:6}(b) show the 
spin-polarized AR contributions, $\la R^{\uarr(\darr)} \ra$, which 
exhibit a detuning in frequency, similarly to the results for
single rings, Figs.~\ref{fig:3}(b) and \ref{fig:5}(b).
This relative dephasing between the different spin
components is also manifested in the total AR, $\la R^{\uarr} \ra+\la
R^{\darr} \ra$, shown in Fig. \ref{fig:6}(c) as a peak splitting
in the regime $|\phi_{\rm m}| \ge 2$.
Note, however, that, contrary to the conductance through single
rings, there are no spin-dependent phase shifts at $\phi_{\rm m}=0$.

We finally summarize a semiclassical explanation\cite{diego} 
for the numerically obtained AR profiles. A semiclassical 
approach to the reflection coefficient is based on expressing
the Green function $\cal G$ in Eq.\ (\ref{t-nm}) 
by a sum over contributions from
 classical backscattered paths (starting and ending at the same
lead)\cite{baranger1,reviews}. The quantum transmission and 
reflection coefficients are then semiclassically approximated 
by double sums over products of such paths. The so-called
diagonal approximation consists in considering for the AR only 
pairs of identical backscattered paths and pairs of an orbit 
and its time-reversed partner. In the presence of
a nonuniform field (in the adiabatic regime) the latter orbit pairs contribute with a 
phase factor where the phase is given by 
$4 \pi  \phi_{\rm eff}^{\uarr(\darr)} / \phi_0$ with
$  \phi_{\rm eff}^{\uarr(\darr)}$  defined in Eq.~(\ref{phi-eff}).
In ring geometries the backscattered orbits can be 
organized according to the number $w$ of revolutions
around the center disk. It then can be shown\cite{baranger1} 
that for a classically chaotic geometry backscattered 
paths with winding number $w=0$ lead to the broad 
(lorentzian-type) backgrounds of the weak-localization
profiles in Fig.~\ref{fig:6}. 
By generalizing a corresponding 
expression\cite{tanaka} for orbits with $w \ge 1$ to the 
case with Berry phases due to $B_{\rm i}$ one finds as a semiclassical
approximation
to the oscillatory part of the AR for small $\phi_{\rm m}$\cite{diego}:
\be
\label{semicl}
\langle \delta R^{\uarr(\darr)}(\phi_{\rm m}) \rangle_{\rm sc}
\simeq \sum_{w = 1}^\infty \ \exp{[-\beta(\phi_{\rm m}) w]} \ \cos\left(4\pi w
     \phi_{\rm m}/\phi_0 \mp 2\pi w \cos \alp \right) \; .
\ee
Here, the exponent $\beta$ involves a quadratic dependence
on $\phi_{\rm m}$ and further depends on classical properties
of the chaotic cavity, i.e.\ the classical escape rate
and the variance of winding number distributions\cite{tanaka}.
In Eq.~(\ref{semicl}) 
we further assumed a fixed (mean) Berry phase for a given 
winding number and neglected its weak influence on $\beta$.

Eq.~(\ref{semicl}), which relies on the above mentioned 
diagonal approximation, can explain the main qualitative features
of the numerically obtained AR oscillations in Fig.~\ref{fig:6}(b,c),
namely the absence of a spin-dependent shift at $B_0=0$ and
the dephasing of the oscillations associated with different
spin. We stress, however, that the semiclassical approach
presented here remains incomplete, since it does not give the correct
amplitudes. A more detailed discussion of Eq.~(\ref{semicl})
and its present limitations will be given in Ref.~\cite{diego}.
For further semiclassical treatments invovling Berry phases
see e.g.\ Refs.~\cite{LF91,BK98}.

%
%%%%%%%%%%%%%%%%%%%%%%%%%%%%%%%%%%%%%%%%%%%%%%%%%%%%%%%%%%%%%%%%%%%%
%%%%%%                  NEW SECTION
%%%%%%%%%%%%%%%%%%%%%%%%%%%%%%%%%%%%%%%%%%%%%%%%%%%%%%%%%%%%%%%%%%%%
%
\section{Concluding remarks}
Recent experimental progress in generating nonuniform
magnetic fields on micron scales in ballistic phase-coherent conductors
has partly motivated this work on geometrical-phase effects on
quantum transport. The observation of a Berry phase requires adiabaticity. 
We thus have generalized adiabaticity criteria
to two-dimensional ballistic ring geometries and showed that the
frequencies of both, angular and radial motion, have to be small compared
to the Larmor frequency of the electron spins. 
However, even if the adiabaticity conditions are not met, 
generalizations of the geometrical phase, such as Aharonov-Anandan 
phases\cite{aharonov3} will pertain\cite{BK2}.
The corresponding theoretical treatment amounts to replace the tilt angle  
$\alp$ by an effective angle $0<\alp'<\alp$\cite{stern2}. 
An application to 1d quantum transport shows distinct effects of such phases 
on the conductance in the nonadiabatic regime\cite{martina}.

Assuming adiabaticity we then showed that the spin-dependent
 magneto conductivity of single ring geometries as well as 
the averaged resistance of ensembles of asymmetric rings
clearly exhibits the influence of geometrical phases which should
be observable in corresponding transport experiments. 

Magneto oscillations in the weak-localization profile and
their dependence on the geometrical phase could be qualitatively
explained using a semiclassical formula for the averaged
reflection based on the so-called diagonal approximation.
However, a quantitative semiclassical theory for weak
localization is still lacking.
It presumably requires the consideration of 
off-diagonal path contributions\cite{SR00} to the average resistance
that are not negligible.
Their computation remains as an open question in semiclassical 
quantum transport and as a challenge for future theoretical work
in this direction.

%
%%%%%%%%%%%%%%%%%%%%%%%%%%%%%%%%%%%%%%%%%%%%%%%%%%%%%%%%%%%%%%%%%%%%
% 
\acknowledgments
We are grateful to M.\ Hentschel, R.A.\ Jalabert, 
P.E.\ Lindelof, S.\ Pedersen, and D.\ Weiss for helpful discussions.
%
%
%%%%%%%%%%%%%%%%%%%%%%%%%%%%%%%%%%%%%%%%%%%%%%%%%%%%%%%%%%%%%%%%%%%%
%

%
%%%%%%%%%%%%%%%%%%%%%%%%%%%%%%%%%%%%%%%%%%%%%%%%%%%%%%

%%%%%%%%%%%%%%% Figures  %%%%%%%%%%%%%%%%%%%%%%%%%%%%%

%%%%%%%%%%%%%%%%%%%%%%%%%%%%%%%%%%%%%%%%%%%%%%%%%%%%%%

% \newpage
% \vspace*{5cm}

\vspace*{2cm}

%%%%%%%%%%%%%%%%%%%%%%%%%%%%%%%%%%%%%%%%%%%%%%%%%%%%%%

\vspace*{1cm}

%%%%%%%%%%%%%%%%%%%%%%%%%%%%%%%%%%%%%%%%%%%%%%%%%%%%%%

\vspace*{1cm}

%%%%%%%%%%%%%%%%%%%%%%%%%%%%%%%%%%%%%%%%%%%%%%%%%%%%%%

\vspace*{1cm}

%%%%%%%%%%%%%%%%%%%%%%%%%%%%%%%%%%%%%%%%%%%%%%%%%%%%%%

\vspace*{1cm}

%%%%%%%%%%%%%%%%%%%%%%%%%%%%%%%%%%%%%%%%%%%%%%%%%%%%%%

%
%\end{multicols}
%


\begin{references}
% {10}

\bibitem{ferry} D.K.\ Ferry and S.M.\ Goodnick, {\em Transport in Nanostructures}
(Cambridge University Press, Cambridge, 1997).

\bibitem{aharonov} Y. Aharonov and D. Bohm, Phys. Rev. {\bf 115}, 485 (1959).

\bibitem{berry} M. Berry, Proc. R. Soc. London A {\bf 392}, 45 (1984).

\bibitem{loss} D. Loss, P. Goldbart, and A.V. Balatsky, Phys. Rev. Lett. {\bf
65}, 1655 (1990); see also: D. Loss and P. Goldbart, Phys. Rev.\ B {\bf 45}, 
13544 (1992).

\bibitem{stern} A. Stern, Phys. Rev. Lett. {\bf 68}, 1022 (1992).

\bibitem{loss2} D. Loss, H.\ Sch\"oller, and P. Goldbart, Phys.\ Rev.\ B {\bf 48}, 
15218 (1993).

\bibitem{kawabata} S.\ Kawabata, Phys. Rev.\ B {\bf 60}, R8457 (1999).

\bibitem{beenakker} S.A.\ van Langen, H.P.A.\ Knops, J.C.J.\ Paasschens,
and C.W.J.\ Beenakker,  Phys. Rev.\ {\bf 59}, 2102 (1999).

\bibitem{loss3} D. Loss, H.\ Sch\"oller, and P. Goldbart,  Phys. Rev.\ B {\bf 59},
13328 (1999); H.A. Engel and D. Loss, preprint cond-math/0002396, to appear in
Phys. Rev.\ B.

\bibitem{spin-orbit} 
Y.\ Meir, Y.\ Gefen, and O.\ Entin-Wohlman, Phys. Rev. Lett. {\bf 63}, 798 (1989);
H.\ Mathur and A.D.\ Stone,  Phys. Rev. Lett. {\bf 68}, 2964 (1992);
A.G.\ Aronov and Y.B.\ Lyanda-Geller, Phys. Rev. Lett. {\bf 70} 343 (1993); 
T.Z.\ Qian and Z.B.\ Su, Phys.~Rev.~Lett. {\bf  72}, 2311 (1994).

\bibitem{weiss} P.D. Ye, S. Tarucha, and D. Weiss, 
in Proc. of the 24th Int.\ Conf.\ on ``The Physics of Semiconductors''
(World Scientific, Singapore, 1998).

\bibitem{morpurgo} A.F.\ Morpurgo, J.P.\ Heidda, T.M.\ Klapwijk, B.J.\ van Wees, 
and G.\ Borghs,  Phys. Rev. Lett. {\bf 80}, 1050 (1998).

\bibitem{grundler} D.~Grundler,  Phys. Rev. Lett. {\bf 84}, 6074 (2000).

\bibitem{baranger1}  
H.U. Baranger, R.A. Jalabert, and A.D. Stone, Chaos {\bf 3}, 665 (1993)
and references therein.

\bibitem{reviews} For recent reviews see e.g.: 
K.~Richter, {\em Semiclassical Theory of Mesoscopic Quantum Systems} (Springer, Berlin, 2000); 
R.A.\ Jalabert, in {\em New Directions in Quantum Chaos}, ed. by
G.~Casati, I.~Guarneri, and U.~Smilansky (IOS Press, Amsterdam, 2000).

\bibitem{LF91} R.G. Littlejohn and W.G. Flynn, Phys. Rev. A {\bf 44}, 5239
(1991). 

\bibitem{BK98} J. Bolte and S. Keppeler, Phys. Rev. Lett. {\bf 81}, 1987
(1998); Ann.~Phys.~(NY) {\bf 274}, 125 (1999).

\bibitem{aharonov2} Y. Aharonov, E. Ben-Reuven, S. Popescu, and D. Rohrlich,
Phys. Rev. Lett. {\bf 65}, 3065 (1990).

\bibitem{note5} 
For the cases numericaly studied the expression $\partial \alp/\partial r$
is  a Lorentzian for which the
integral (\ref{integral}) has no analytical solution. However,
 considerations of  different limits yield results which
  coincide with the one of Eq. (\ref{2Dadiab-a}).

\bibitem{lindelof} see, e.g. A.E. Hansen, S. Pedersen, A. Kristensen,
C.B. S{\o}rensen, and P.E. Lindelof, Physica E {\bf 7}, 776 (2000).

\bibitem{geim} S.V. Dubonos, A.K. Geim, K.S. Novoselov, J.G.S. Lok, J.C. Maan,
and N. Henini, Physica E {\bf 6}, 746 (2000).

\bibitem{experiment} 
Corresponding transport experiments are in progress (D.~Weiss, private communication).

\bibitem{landauer} R.~Landauer, IBM J.~Res.~Dev.~{\bf 1}, 233 (1957);
for reviews see Ref.~\cite{ferry} or S.~Datta, {\em Electronic
Transport in Mesoscopic Systems} (Cambridge University Press, Cambridge, 1997).
% M. B\"uttiker, Y. Imry, R. Landauer, and S. Pinhas, Phys. Rev B {\bf 31}, 6207
% (1985); Y. Imry, contribution to {\it Directions in Condensed Matter Physics},
% ed. by G. Grinstein and E. Mazenko, World Publishing, Singapore, (1986). 

\bibitem{fisher} D.S. Fisher and P.A. Lee, Phys. Rev. B {\bf 23}, 6851 (1981).

% \bibitem{economou} E.N. Economou, {\it Green's Functions in Quantum Physics},
% Springer-Verlag (1990).

\bibitem{ensslin} S. Brosig, K. Ensslin, A.G. Jansen, C. Nguyen, B. Brar,
M. Thomas, and H. Kroemer, Phys. Rev. B {\bf 61}, 13045 (2000). 

\bibitem{pichugin} K.N. Pichugin and A.F. Sadreev, Phys. Rev. B {\bf 56}, 9662 (1997).

\bibitem{imry} M.~B\"uttiker, Y.~Imry, and M.~Y.~Azbel, Phys.~Rev.~A {\bf 30}, 1982 (1984).

%\bibitem{wees} B.J. van Wees, L.P. Kouwenhoven, H. van Houten,
%C.W.J. Beenakker, J.E. Mooij, C.T. Foxon and J.J. Harris, Phys. Rev. B {\bf 38}, 3625 (1988).

% \bibitem{note1} A $\phi_0/2$ component can eventualy appears, but the pattern
% will still be $\phi_0$-periodic. However, for a large amount of flux
% $\phi_{\rm m}$, the fact of having a ring with finite width gives rise to
% distorsions in the shape of the oscillations (see figure...(b) and \cite{pichugin}). 

% \bibitem{note2} The zero-field phase of the magnetoconductance oscillations is
% rather sensitive to the electronic energy due to the conductance fluctuations
% at low temperatures \cite{lindelof}. The presence of an effective texture
% $V_{\rm eff}$ would affect the final result, contributing to the phase shift
% together with the geometricaly generated one. This problem desapears in the
% case of energy-averaged magnetoconductance. 

\bibitem{note3} A refined treatment of the dephasing beyond 
the simple picture based on sine functions can be achieved by extending the 
approach of Ref.\ \cite{imry} to the case of geometrical phases\cite{martina}.

\bibitem{martina} M. Hentschel, D. Frustaglia, and K. Richter, unpublished.

\bibitem{buttiker} M. B\"uttiker and Y. Imry, J. Phys. C {\bf 18}, L467
(1985); M. B\"uttiker,  Phys. Rev. Lett. {\bf 57}, 1761 (1986).

\bibitem{baranger2} H.U. Baranger and A.D. Stone Phys. Rev. B {\bf 40}, 8169
(1989).

\bibitem{note4} 
We see from Eqs.~(\ref{Ag}) and (\ref{Aeff}) that for $r\neq 0 $,
$B_{\rm eff}^{z \ \uarr(\darr)}=B_0 \mp  \phi_0/(4\pi \ r) \
(\partial \cos \alp (r) /\partial r ) $,
where $\partial \cos \alp (r)/ \partial r= - (B_0/B^2) \ \partial B/\partial r$, giving
rise to the anti-symmetry of $B_{\rm eff}^{z \ \uarr(\darr)}$ under the inversion
of $B_0$. 
This holds true also for $r=0$ (where  $A_{g}^{\uarr (\darr)}$ diverges),
since we find for any path $\Gamma$ that
$ (c/e) \oint_{\Gamma} \vec{A}_{g}^{\uarr(\darr)}(B_0) \cdot \vec{{\rm d}l} = - 
(c/e) \oint_{\Gamma} \vec{A}_{g}^{\uarr(\darr)}(-B_0) \cdot \vec{{\rm d}l} + \phi_0 $,
where the last term  is equivalent to a trivial shift of
$2\pi$ in the dynamical phase of the electrons according to the Aharonov-Bohm
theorem\cite{aharonov}. 

\bibitem{spivak} B.L.\ Altshuler, A.G.\ Aronov, and B.Z.\ Spivak,
Pi'sma Zh.\ Eksp.\ Teor.\ Fiz.\ {\bf 33}, 101 (1981) [JETP Lett.\ {\bf 33},
94 (1981)].

\bibitem{tanaka} S.~Kawabata and K.~Nakamura, Phys.\ Rev.\ B {\bf 57}, 6282 (1998).

\bibitem{weiss2} P.D.\ Ye and D. Weiss (private communication).

\bibitem{diego} D. Frustaglia and K. Richter, in preparation.

\bibitem{aharonov3} Y. Aharonov and J. Anandan, Phys. Rev. Lett. {\bf 58}, 1593 (1987).

\bibitem{BK2} A corresponding semiclassical approximation\cite{BK98} is also not
              restricted to the adiabatic limit.

\bibitem{stern2} A. Stern, in
% {\it Geometric Phases in Mesoscopic Systems - From
% the Aharonov-Bohm Effect to Berry Phases}, contribution to 
{\it Mesoscopic Electron Transport}, 
ed.\ by L.L. Sohn el al.\ (Kluwer Academic Publishers, New York, 1997).

\bibitem{SR00} M.~Sieber and K. Richter, preprint MPI-PKS/0008004,
                submitted to Physica Scripta, 2000.

\end{references}
\end{document}